\documentclass{article}
\usepackage{graphicx}
\usepackage{epsf}
\usepackage{psfig}
\begin{document}
\bigskip
\centerline{QUASI-INTEGRABILITY IN A CLASS OF SYSTEMS}
\centerline{GENERALIZING THE PROBLEM OF TWO FIXED CENTERS}
\vskip1.5cm
\centerline{ A. ALBOUY}
\smallskip
\centerline{\it IMCCE}
\centerline{\it 77, av. Denfert-Rochereau, 75014 Paris, France}
\centerline{\it albouy@bdl.fr}

\bigskip

\centerline{ T.J. STUCHI}
\smallskip
\centerline{\it Instituto de F\'{\i}sica, Universidade Federal do Rio de
Janeiro}
\centerline{\it Caixa Postal 68528}
\centerline{\it 21941-972, Rio de Janeiro, Brazil}
\centerline{\it tstuchi@if.ufrj.br}

\bigskip

\noindent{\bf Abstract.} The problem of two fixed centers is a classical
integrable
problem,
stated and integrated by Euler in 1760. The integrability
is due to the unexpected first integral $G$. Some straightforward
generalizations of the problem
still have the generalization of $G$ as a first integral, but
do not
possess the energy integral. We present some numerical integrations
suggesting that in the domain of bounded orbits the behavior of these
{\it a priori} non hamiltonian systems is very similar to the behavior
of usual quasi-integrable systems.

\bigskip

\noindent{\bf The equations.}
Euler's problem in the plane (see  Figure 1) is defined by the system of
differential equations
$$\ddot x=-a(x_A,y)x_A-b(x_B,y)x_B,\quad \ddot y=-a(x_A,y)y-b(x_B,y)y.\eqno(1)$$
The two fixed centers are the points $(1,0)$ and $(-1,0)$, and the
moving particle is the point $(x,y)$. We have set $x_A=x-1$, $x_B=x+1$,
$$a(\xi,\eta)=m_A(\xi^2+\eta^2)^{-3/2},\qquad
b(\xi,\eta)=m_B(\xi^2+\eta^2)^{-3/2}.\eqno(2)$$
The problem can be defined in the 3-dimensional space in the same way, and
is also integrable,
as was noticed by Euler. However, we will restrict ourselves to the
planar case.

The first step in Euler's integration was to exhibit two independent
first integrals of the
motion. One is the energy $$H=(\dot x^2+\dot y ^2)/2-\tilde a
(x_A,y)-\tilde b(x_B,y),$$
with $\tilde a(\xi,\eta)=m_A(\xi^2+\eta^2)^{-1/2}$, $\tilde b(\xi,\eta)
=m_B(\xi^2+\eta^2)^{-1/2}$. We will
call the second one Euler's integral:
$$G=C_AC_B-2\tilde a(x_A,y)x_A+2\tilde b(x_B,y)x_B,$$
with $C_A=x_A\dot y-y\dot x$, $C_B=x_B\dot y-y\dot x$.
Euler continued the integration, eliminating
the second derivatives in (1) using the first integrals, and separating
the variables.

Our generalization is simply to consider system (1) in the case where $a$
and $b$ are any homogeneous functions of degree $-3$. Indeed, we want to put
a little restriction on these
homogeneous functions. We will suppose
that both differential forms $\xi a(\xi,\eta) (\xi d\eta-\eta d\xi)$ and
$\eta a(\xi,\eta) (\xi d\eta-\eta d\xi)$ are exact forms on the plane minus
the origin, and that the same is true when we change $a$ in $b$. This
hypothesis comes from the study of the problem of one fixed center (see [2] and
[3]). It
is not a strong restriction: the forms are already closed, so the condition
on each function is just the cancellation of two scalar quantities, namely
the integrals of both forms on a closed path around the origin.

It can be shown that any function $a(\xi,\eta)$ satisfying the above conditions
comes from a function $A(\xi,\eta)$ homogeneous of degree $1$ as follows. Let us
denote by $A_\xi$, $A_\eta$ the first derivatives of the function $A$ and by
$A_{\xi\xi}$, $A_{\xi\eta}$, $A_{\eta\eta}$ the second derivatives.
Then
$$a(\xi,\eta)=\eta^{-2}A_{\xi\xi}=-\xi^{-1}\eta^{-1}A_{\xi\eta}=\xi^{-2}
A_{\eta\eta}.$$
The function $a(\xi,\eta)$ in  Euler's case $(2)$ is obtained in this way from
the function $A=m_A(\xi^2+\eta^2)^{1/2}$. We also have $A_\xi=\xi\tilde a$.
It was discovered by the first author that $G$ persists in the form
$$G=C_AC_B-2A_\xi(x_A,y)+2B_\xi(x_B,y),$$
where $B(\xi,\eta)$ and $B_\xi(\xi,\eta)$ are associated to the function
$b$ in the same way as $A$ and $A_\xi$ are associated to $a$.
In general, no integral takes the place
of the energy integral.

\bigskip
\noindent
{\bf Quasi-integrability.} We report our numerical exploration of these
generalized Euler's
problems, showing three examples that seem to us significant. In all cases
we met, the
result is either escape or quasi-integrable behavior. The third experiment
displays some
islands suggesting non integrability. Magnifying  the neighborhood
of a saddle point, a domain of irregular dynamics can be observed.

The obvious choice for a Poincar\'e section is to fix the integral $G$ and take,
for example, $y=0$ $(\dot y >0)$.
In each case, we show the iterates of some points of  this Poincar\'e
mapping, the central orbit
in the section and some typical quasiperiodic orbit.
All the orbits in a given field of forces have the
same value of Euler's integral. Since the examples have very large orbits,
we have taken throughout the numerical experiments a somewhat arbitrary
cut-off criterion given by the value of any coordinate or velocity greater than
a thousand.
In all the examples, we have taken $A(\xi,\eta)=(5\xi^2-\xi\eta+5\eta^2)/10r$,
where $r=\sqrt{\xi^2+\eta^2}$, which corresponds to $a(\xi,\eta)=(5\xi^2+3\xi
\eta+5\eta^2)/10r$.
\bigskip

\noindent{\bf First example.} Figures 2 to 4 correspond to
$B(\xi,\eta)=-(\xi^3-3\xi \eta^2)/4r^2+r$, and thus $b(\xi,\eta)=-2(3\xi
\eta^2-\xi^3)/r^6+r^{-3}$.
The Poincar\'e map is close to a linear map, on a whole domain delimited
by the escape
criterion. This is quite strange. An explanation for this phenomena comes
from geometrical
considerations. We have chosen the plane as the domain for the motion, but there
is a natural
bigger domain for this kind of systems. It is the manifold of half lines
drawn from the
origin in a $3-$dimensional vector space. Our plane is from this point of
view just one half
of the natural domain,
a hemisphere chosen arbitrarily. Escaping orbits
appear as
orbits cut by the boundary of the hemisphere (in the classical Kepler
problem, hyperbolas
appear in the same way as cut ellipses). The theoretical grounds for this
remark may be found in
[4].
\bigskip

\noindent{\bf Second example.} In Figures 5 to 7, we have chosen
$B=4(\xi^4+\eta^4)^{1/4}$,
$b=12\xi^2\eta^2(\xi^4+\eta^4)^{-7/4}$. Here the section displays a wide
domain with strong
torsion but we are still very close from an integrable system. This rises
 the question:
what are the integrable systems nearby? We know very few cases where our
generalized
Euler problem is integrable, namely the classical case and its projective
transformations
defined in [4] (which correspond for example to replace $\xi^2+\eta^2$ in
Eq. (2) for $b$ by
any homogeneous quadratic expression in $\xi$, $\eta$, and leave $a$ as it
is.) Because we
needed to get sufficiently many bounded orbits we were probably forced to
stay close  from
integrable cases.

\bigskip

\noindent{\bf Third example and final comments.} In Figures 8 to 11,
we have chosen $B=(3\xi^2-\xi\eta+3\eta^2)/3r$,
$b=(\xi^2+\xi\eta+\eta^2)/r^5$.
Here the system behaves as a typical conservative system close to
an integrable system. We are more accustomed to observe this
in the class of hamiltonian systems, and one can argue that maybe the system is
hamiltonian for some symplectic form. We do not believe so, and rather relate
this quasi-integrability to KAM theory applied to reversible systems
(see  [5], Theorem 2.9).
Our systems are clearly reversible.
\bigskip

\noindent
{\bf Acknowledgments.} We thank Prof. H. Cabral and the  Departamento de
Matem\'atica
da Universidade Federal de Pernambuco for their hospitality, Prof. A.
Neishtadt for
stimulating comments, and Prof. Carles Sim\'o for his rk78 code and his
very useful comments.
\bigskip

\noindent
{\bf References}

[1] L. Euler, {\sl De motu corporis ad duo centra virium fixa attracti},
{\sl Probl\`eme. Un corps \'etant attir\'e en raison r\'eciproque quarr\'ee
des distances vers deux points fixes donn\'es trouver les cas o\`u la
courbe d\'ecrite
par ce corps sera alg\'ebrique.} Opera Omnia, S. 2, vol 6. (1764, 1765,
1760) pp. 209--246,
247--273, 274--293

[2] A. Albouy, {\sl Lectures on the two-body problem}, Classical and
Celestial Mechanics:
The Recife Lectures, H.
Cabral F. Diacu editors, volume ``in press'',
Princeton University Press

[3] G. Darboux, {\sl Sur une loi particuli\`ere de la
force signal\'ee par
Jacobi}, Note 11 au {\sl cours de m\'ecanique}
de Despeyrous, tome premier, A. Hermann, Paris (1884)

[4] P. Appell, {\sl Sur les lois de forces centrales faisant d\'ecrire \`a leur
point d'application une conique quelles que soient les conditions initiales},
American Journal of Mathematics 13 (1891) pp.\ 153--158

[5] J. Moser, {\sl Stable and Random Motions in Dynamical Systems, With Special
Emphasis on Celestial Mechanics}, Princeton University Press (1973)


\begin{figure}[ht]
\centerline{\includegraphics [width=100mm] {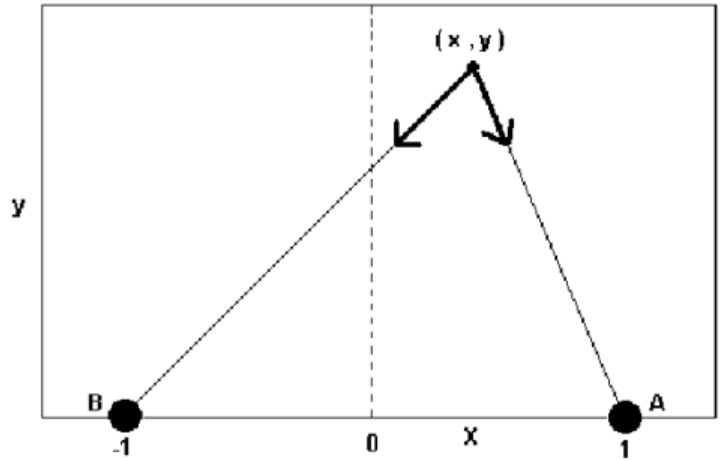}}
\caption{In Euler problem as well as in the generalizations we present,
a particle at the position $(x,y)$ evolves under the action of
two centers A and B, with respective coordinates $(1,0)$ and $(-1,0)$.}
\end{figure}

\begin{figure}[ht]
\centerline{\includegraphics [width=100mm] {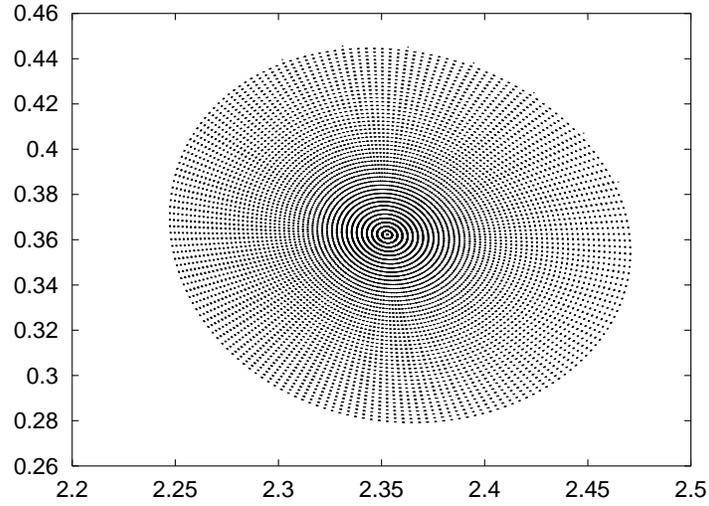}}
\caption{Example 1; $B=-(\xi^3-3\xi\eta^2)/4r^2+r$; $G= 5.50433086$.
Poincar\'e section  $(y=0,\dot y>0)$,
from the central periodic orbit to
the neighborhood of the last  torus before cut-off.}
\end{figure}

\begin{figure}[ht]
\centerline{\includegraphics [width=100mm] {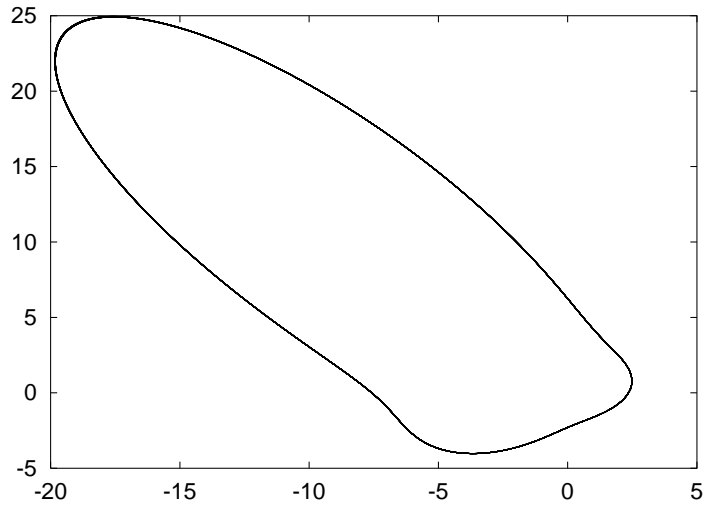}}
\caption{Example 1 again. Configuration space of the periodic orbit:
 $x=2.352375$, $\dot x=0.3621675$ and $\dot y=1.050626$.}
\end{figure}

\begin{figure}[ht]
\centerline{\includegraphics [width=100mm] {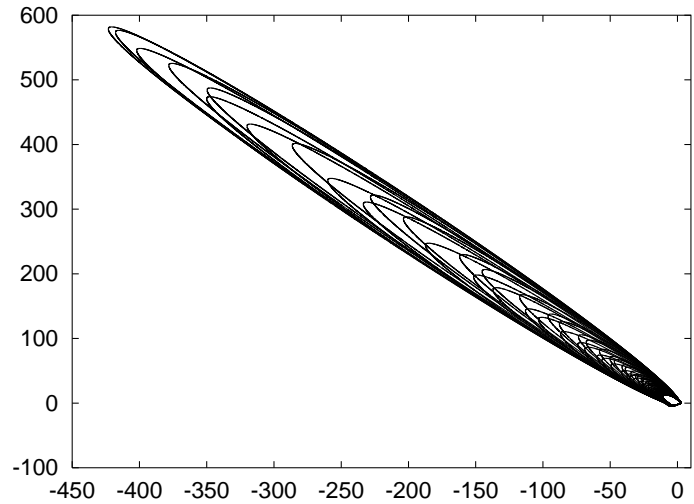}}
\caption {Example 1. The configuration space of the last orbit shown in the Poincar\'e section just
before our escape
criterion is satisfied: $x =3.469875$, $\dot x=0.3621675$ and $\dot
y=0.99058834$.}
\end{figure}

\begin{figure}[ht]
\centerline{\includegraphics [width=100mm]{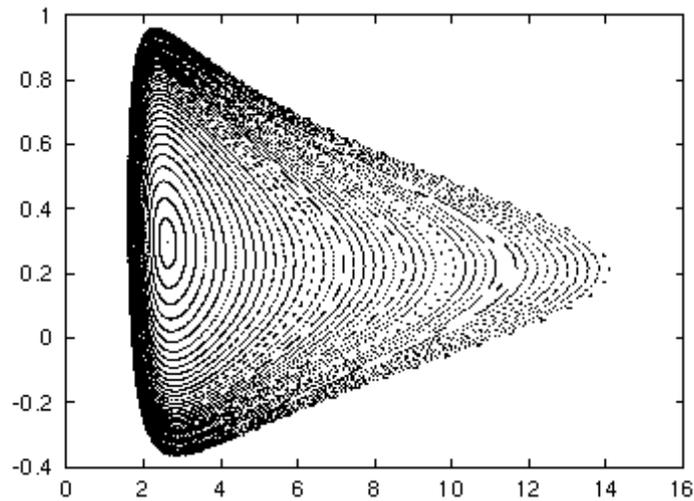}}
\caption {Example 2; $B=-(\xi^3-3\xi\eta^2)/4r^2+r$; $G=10.9200094$.
Same Poincar\'e section as Figure 2, showing again
a phase space
foliated by invariant tori up to the escape orbit.}
\end{figure}

\begin{figure}[ht]
\centerline{\includegraphics[width=100mm] {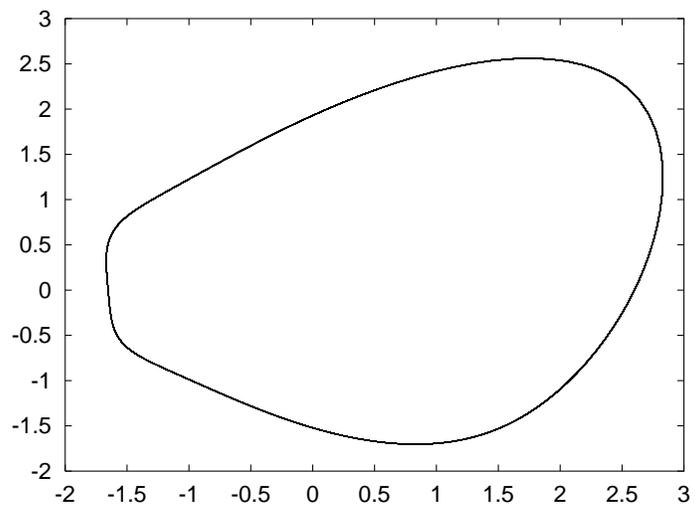}}
\caption{Example 2. The configuration space of the periodic orbit with initial conditions
$x= 2.6$, $\dot x= 0.2934$ and $\dot y=0.8249589$.}
\end{figure}

\begin{figure}[ht]
\centerline{\includegraphics[width=100mm]{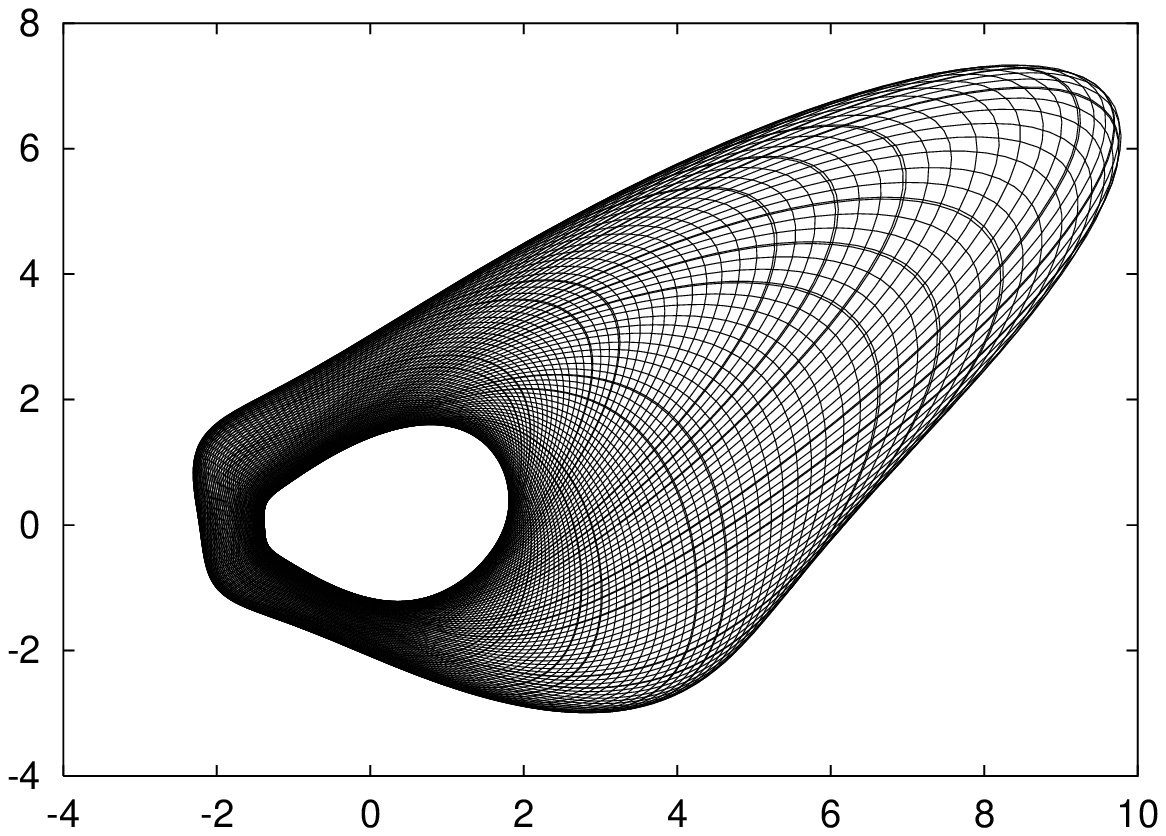}}
\centerline{\includegraphics[width=100mm] {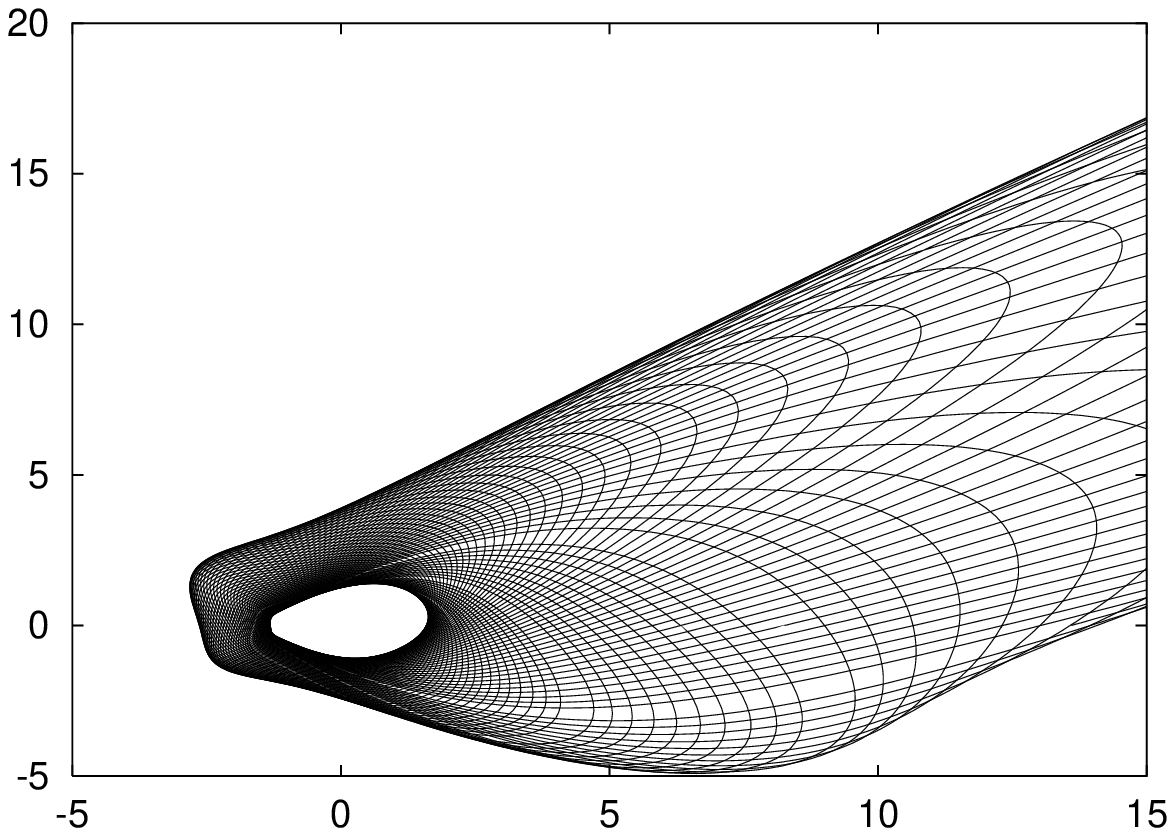}}
\caption{Example 2. Top: A typical torus at $x=6.1$, $\dot x=0.2934$ and
$\dot y=0.3290253$; Bottom: A detail of the  last torus before escape  with
initial conditions $x=13.6$, $\dot x =0.2934$ and $\dot y=0.145976133$.}
\end{figure}
\bigskip

\begin{figure}[ht]
\centerline{\includegraphics [width=100mm]{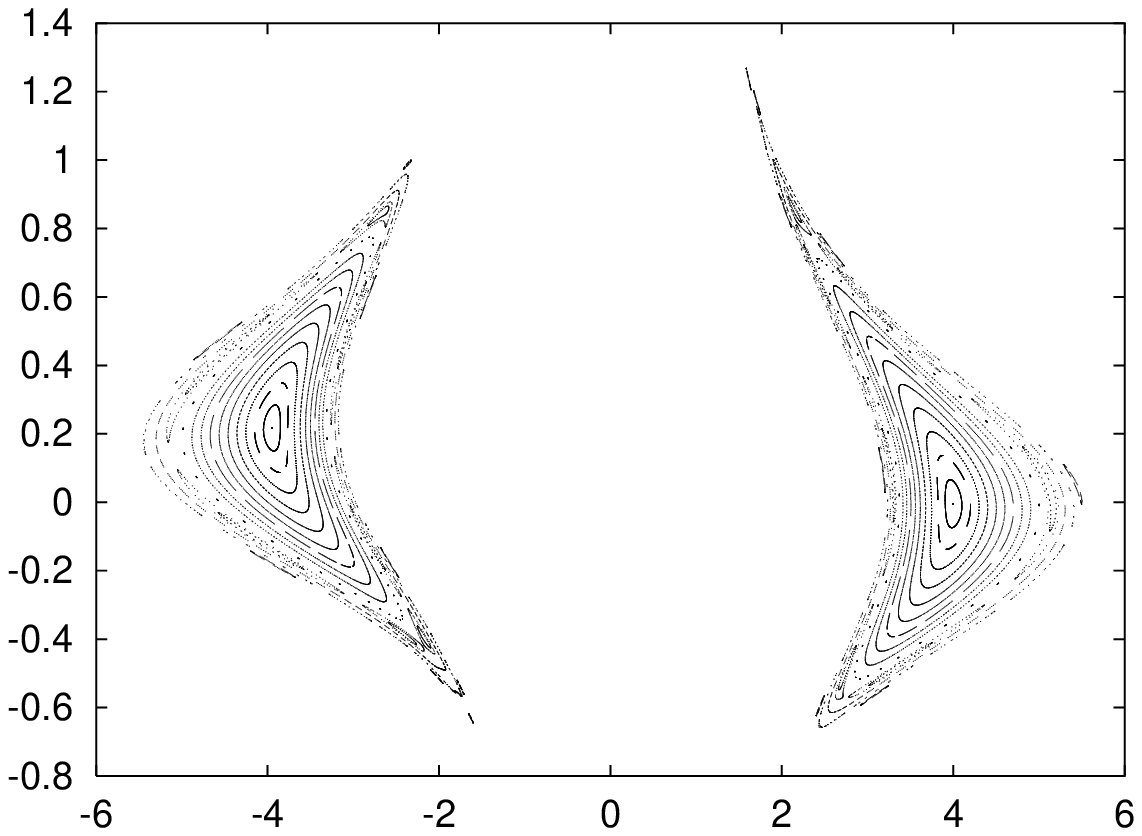}}
\centerline{\includegraphics [width=100mm]{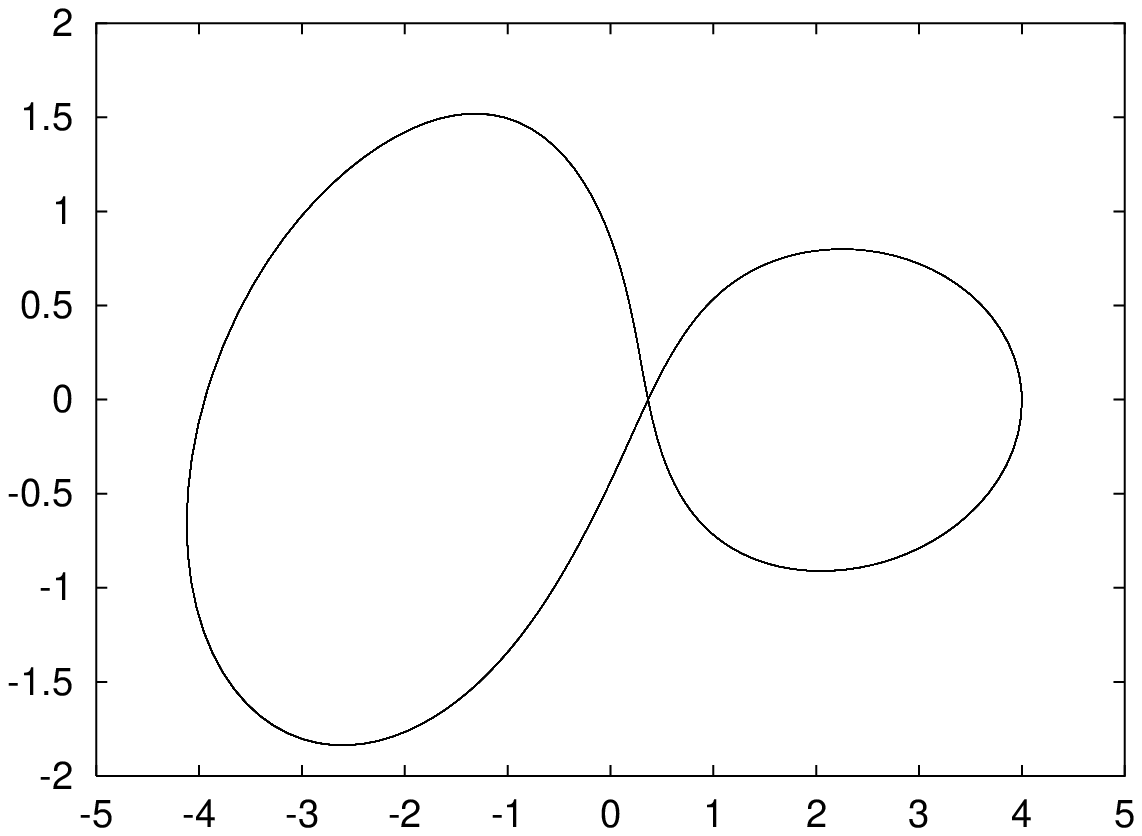}}
\caption{Example 3; $B=(3\xi^2-\xi\eta+3\eta^2)/3r$; $G=1.55230255$. Top:
Poincar\'e section as in Figure 2 starting at the
central periodic orbit up to escaping orbits;
Bottom: The configuration space of the central periodic
orbit  $x=4.0005$, $\dot x=0.191860239$ and $\dot y=0.191860239$.}
\end{figure}

\begin{figure}[ht]
\centerline{\includegraphics[width=100mm]{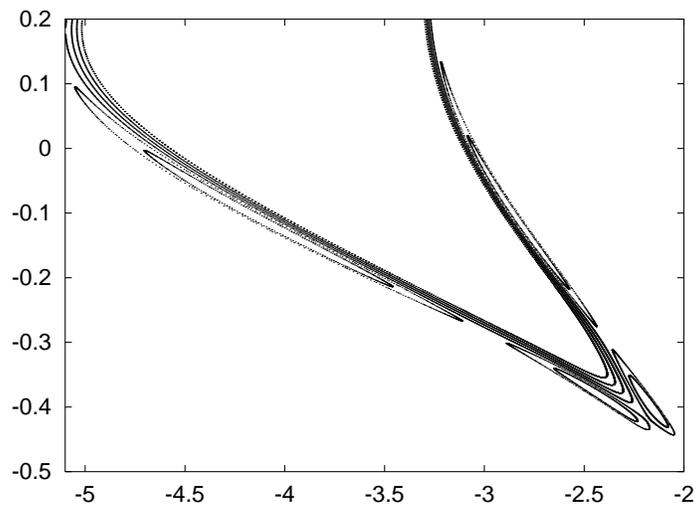}}
\caption{A magnification of Example 3 Poincar\'e Section showing four
islands of a chain with fourteen islands.}

\end{figure}
\begin{figure}[ht]
\centerline{\includegraphics[width=100mm]{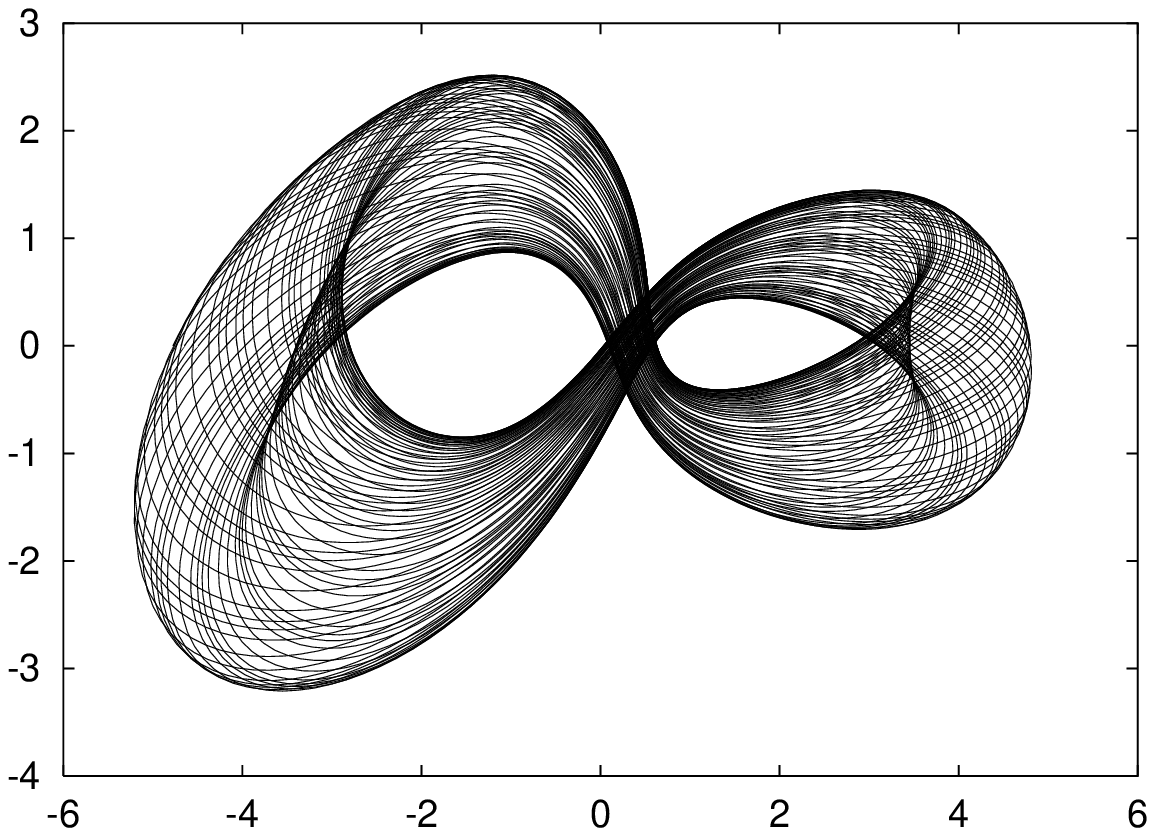}}
\centerline{\includegraphics[width=100mm]{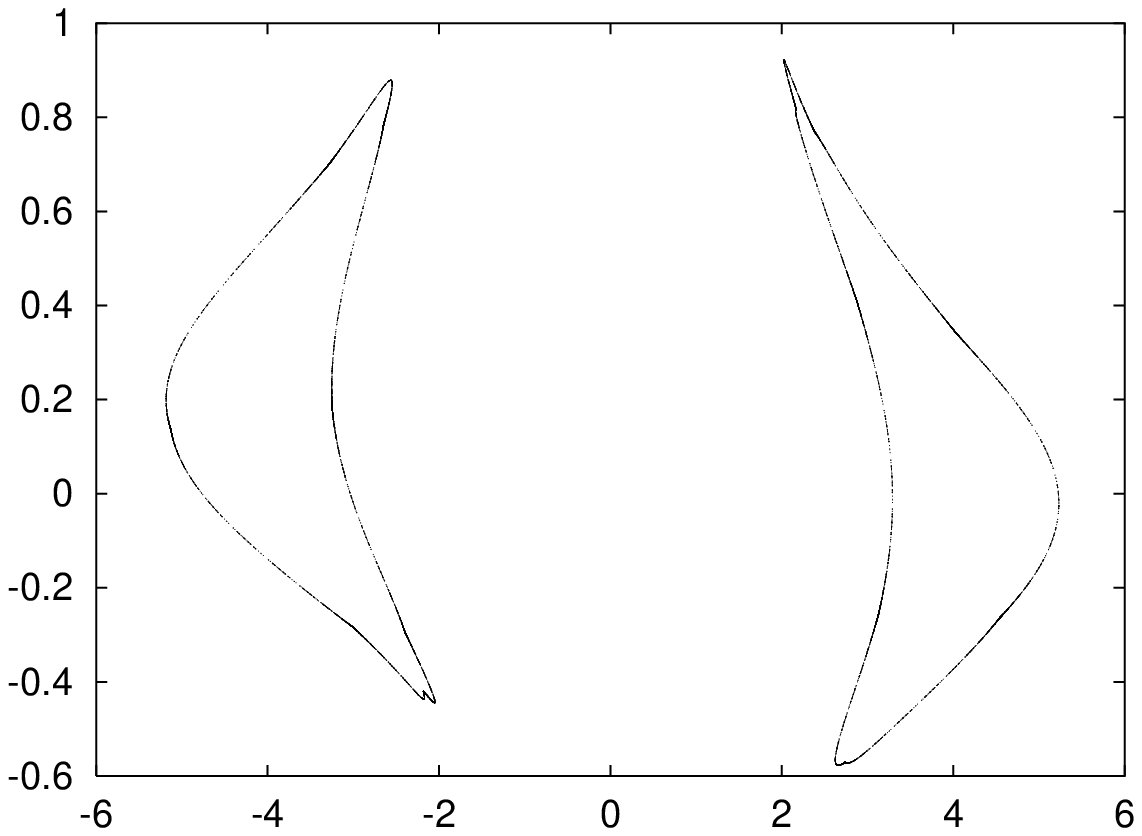}}
\caption{A torus very near the separatrix and its section for Example 3.}
\end{figure}

\begin{figure}[ht]
\centerline{\includegraphics[width=100mm]{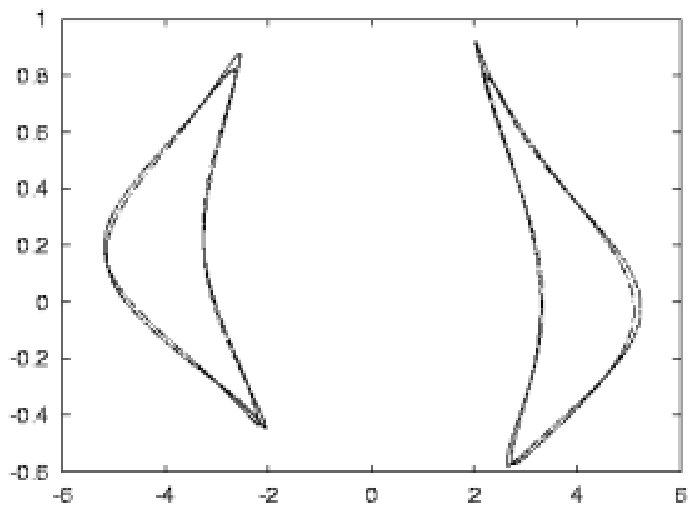}}
\centerline{\includegraphics[width=100mm]{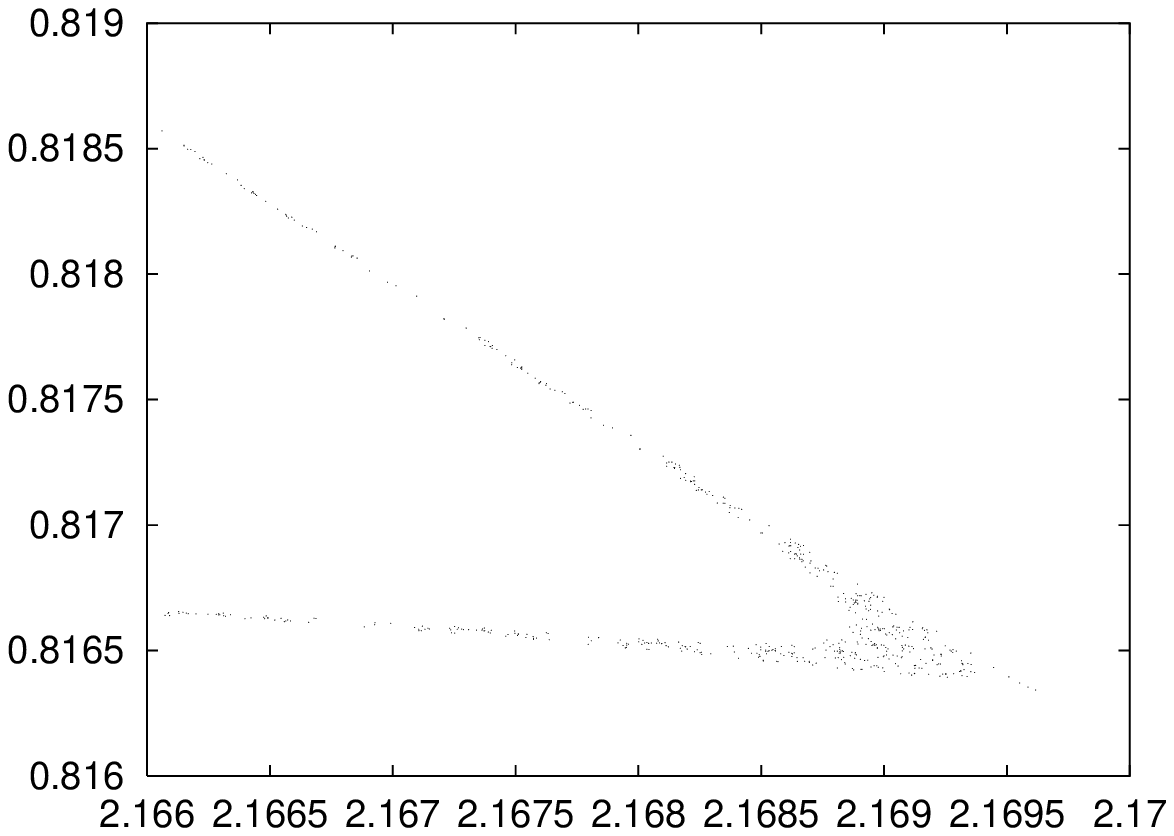}}
\caption{Example 3.  Top: The neighborhood of the islands of the Poincar\'e
section;
Bottom: Its stochastic zones indicating the absence of integrability. }
\end{figure}

\end{document}